\documentclass[prb,12pt,aps]{revtex4-2}

\usepackage{amssymb, amsmath, lmodern}
\usepackage{verbatim}   % useful for program listings
\usepackage{xcolor}      % use if color is used in text
\usepackage{hyperref}   % use for hypertext links, including those to external documents and URLs
\usepackage{setspace} 
\usepackage{graphicx}
\usepackage{subfig}
\usepackage{caption}
\begin{document}
\title{MECHANOSENSITIVE ION CHANNELS: OLD BUT NEW
}
\title{MECHANOSENSITIVE ION CHANNELS: OLD BUT NEW}

\author{U\u{g}ur \c{C}etiner$^{1}$}  
\email{cetiner.ugur@gmail.com}
\affiliation{%
$^1$ Department of Systems Biology, Harvard Medical School, Boston, MA 02115, USA.
}
\maketitle
\section*{Abstract}
 Ion channels orchestrate the communication between cells and their environment. These are special proteins capable of changing their shape to allow ions to pass through membranes in response to stimuli like membrane tension. As the letters are written and exchanged in the form of ionic currents, the energetic cost of this communication is of interest to many. To this end, we introduce the nanoscale thermodynamics of mechanosensitive ion channels, which amounts to defining work and heat for traces obtained during patch-clamp experiments. We also discuss the interplay between ion channel physics and evolution by showing how information-processing capabilities are coupled with the energy landscapes of channels.
%%%%%%%%%%%%%%%%%%%%%%%%%%%%%%%%%%%%%%%%%%%%%%%%%%%%%%%%%%%%%%%%%%%%%%%%%%%%%%%%%%%%%%%%%%%%%%%%%%%%%%%%%%%%%%%%%%%%%%%%%%%%%%%%%%%%%%%%%%%%%%%%%%%%%%%%%%%%%%%%%%%%%%%%%%%%%%%%%%%%%%%%%%%%%%%%%%%%%%%%%%%%%%%%%%%%%%%%%%%%%%%%%%%%%%%%%%%%%%%%%%%%%%%%%%%%%%%%%%%%%%%%%%%%%%%%%%%%%%%%%%%%%%%%%%%%%%%%%%%%%%%%%%%%%%%%%%%%%%%%%%%%%%%
 \section*{Introduction}
Cells are surrounded by membranes that separate their internal world from the outside. Without a membrane, life as we know it could not maintain the necessary internal conditions for biochemical reactions, energy production, and homeostasis. But high concentrations of cellular materials create an osmotic force that pushes water into cells, making them susceptible to osmotic fluctuations. Consequently, cells need to control and regulate their internal water levels. In their membranes, there are force-sensing molecules, which allow passage of charged atoms (ions) and small osmolytes in response to a mechanical stimulus such as membrane tension. These gated pores, known as mechanosensitive (MS) ion channels, are primarily evolved to reduce the internal turgor pressure generated by excessive water influx. They are found in all three domains of life: Bacteria, Archaea, and Eukarya \cite{kung2005possible, martinac2004mechanosensitive}. In higher organisms, mechanosensitive ion channels play key roles in sensing touch, organ distension, proprioception, cardiovascular regulation, balance and hearing \cite{haswell2011mechanosensitive,sukharev2022mechanosensitive}. Bacterial MS channels are the most well-characterized ones because they were discovered first. We can consider them as the ``hydrogen atoms" of mechanosensation. By studying them, we get a clear view of how these tiny molecular machines work in all sorts of organisms. In this article, we will focus on their gating principles from a modern perspective and explore how ion channel physics ties into evolution.
  
\begin{figure}
    \centering
    \includegraphics[width=\linewidth]{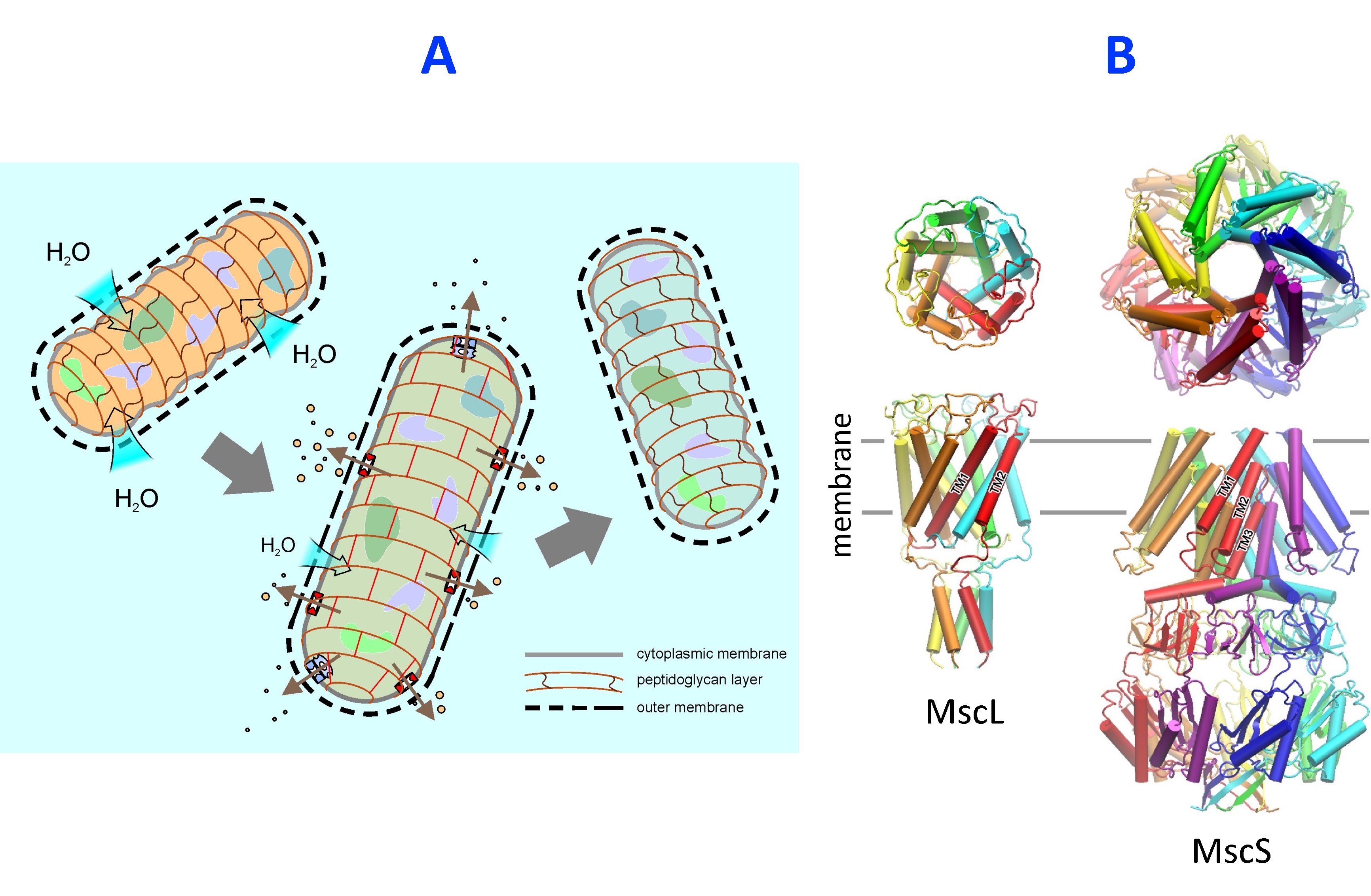}
    \caption{\textbf{A} Bacterial response to osmotic downshock. Elastic deformation of the cell wall and stretching of the cytoplasmic (inner) membrane accompany water influx. Mechanosensitive channels open and release small osmolytes in response to an increase in the membrane tension. When the tension and volume return to normal, the channels close. \textbf{B} Crystal structures of the mechanosensitive channel proteins MscL and MscS. Mycobacterium tuberculosis MscL crystal structure at 3.5 \r{A} (2OAR.pdb) \cite{chang1998structure}. The channel opens at tensions approaching the lytic threshold of membranes ($10-14 mN/m$) and is classified as an emergency valve. Escherichia coli MscS crystal structure at 3.9 \r{A} (2OAU.pdb) \cite{bass2002crystal}. We provide the top and side views for both channels. Figures are adapted from \cite{cetiner2018nanoscale}.}
    \label{F1}
\end{figure}

Bacteria often face sudden changes in the osmolarity of their environment. To cope with these changes, they use various strategies. Under hyperosmotic conditions, when water leaves the cell and turgor pressure decreases, bacteria produce or take in certain compatible osmolytes to restore the positive turgor pressure needed to keep growing \cite{wood1999osmosensing}. However, when the external osmolarity suddenly drops—such as during rainfall—water rushes into cells. This rapid water influx generates significant turgor pressure and stretches the elastic peptidoglycan layer within milliseconds \cite{ccetiner2017tension}. As the inner membrane fully unfolds, it starts to experience tension. This tension is what triggers MS channels to open (gate), letting out ions and small molecules to reduce the pressure, as depicted in Fig. \ref{F1}A. If there is a sudden and strong osmotic down-shock, the inner membrane can stretch the peptidoglycan layer beyond its limits, potentially causing a break and forming a protrusion that can lead to cell rupture and death \cite{bialecka2015rate}. The rate of the tension generation during a down-shock is determined mainly by the rate of water influx, and the rate of osmolyte efflux via MS channels \cite{levina1999protection,bialecka2015rate,ccetiner2017tension}. If channels can reduce the osmotic gradient quickly enough, the cell can be saved. If not, it may burst.

Most of what we know about how water and solutes move in and out of cells  comes primarily from the bacterial species \textit{E. coli}. Water influx accompanying osmotic down-shock takes place directly through the lipid bilayer and is accelerated by water channels (aquaporins) \cite{calamita1995molecular,calamita2000escherichia,finn2015evolution}. Ions and small osmolytes are mainly released by two types of channels: mechanosensitive channels of small conductance (MscS) and those of large conductance (MscL). The crystal structures of these channels are given in Fig. \ref{F1}B. MscL consists of five identical subunits, each 136 amino acids long. 
It has two transmembrane domains that sense tension. This channel only opens under extreme tension (10-14 $mN/m$), acting as a last resort to prevent cells from bursting by forming a wide, non-selective pore in the membrane, about $30$ \r{A} in diameter \cite{sukharev1994large,sukharev1999energetic}. Unlike MscL, MscS is a heptamer of subunits with three TM domains and requires less tension to open (5-7 $mN/m$). The $C$-terminal ends of subunits form a large hollow cytoplasmic cage domain, which can sense the osmotic pressure. This sensor allows MscS to integrate more information, as the channel, in addition to membrane tension, can modulate its open probability based on cytoplasmic crowding. We discuss this point further in the subsequent sections. 

If both MscS and MscL channels are deleted (double knockout strain), \textit{E. coli} cells become much more sensitive to sudden drops in osmolarity. Re-expression of either of the channels, on the other hand, extends the range of tolerable osmotic downshift from 300 mOsm to nearly 1000 mOsm \cite{levina1999protection,ccetiner2017tension}. To put this in perspective, standard LB growth medium has an osmolarity of nearly 450 mOsm, while distilled water has virtually 0 mOsm. Normally, \textit{E. coli} cells grown in LB have a higher internal osmolarity than the medium to maintain a positive turgor pressure. Therefore, when the cells grown in LB are transferred into distilled water, they suddenly experience an osmotic down-shock of 450 mOsm or more.

Bacteria do not have brains, yet they still need to process information and compute. MS channels can be used to compute if the internal water levels are reaching dangerous levels. An ion channel, which can be considered as a two-state system, is capable of storing and processing one bit of information. In the absence of danger, the ion channel remains closed, which we can arbitrarily code as 0. When there is danger, indicated by high membrane tension, the channel opens, which we can code as 1. To process information and perform computation, a physical system that can exist in different states and an external force that can alter the probability of these states are needed. For MS channels, this force could be the membrane tension, and the open probability can be manipulated with the tension, coupling the state of the system to environmental changes. Our focus here is on the energetics of computations carried out by ion channels. In principle, computation can be performed with no energy cost, but deleting a bit of information requires at least $k_BT \ln(2)$ of heat dissipation into the environment \cite{bennett1973logical,landauer1961irreversibility}. Here $k_B$ is the Boltzmann constant and $T$ is the ambient temperature. The laws of physics impose these limits, but biological systems are not necessarily bound by them. To investigate if biological systems approach these limits in processing information, we need to calculate the heat dissipated into the environment and work done on channels during their gating. Recent advancements in non-equilibrium thermodynamics of small systems allow us to define these thermodynamic quantities at the level of a single stochastic trajectory \cite{seifert2012stochastic,jarzynski2012equalities}. The answers to these questions are closely related to how the gating parameters are shaped by billions of years of evolution and can help us understand how biological systems navigate their parameter space.
%%%%%%%%%%%%%%%%%%%%%%%%%%%%%%%%%%%%%%%%%%%%%%%%%%%%%%%%%%%%%%%%%%%%%%%%%%%%%%%%%%%%%%%%%%%%%%%%%%%%%%%%%%%%%%%%%%%%%%%%%%%%%%%%%%%%%%%%%%%%%%%%%%%%%%%%%%%%%%%%%%%%%%%%%%%%%%%%%%%%%%%%%%%%%%%%%%%%%%%%%%%%%%%%%%%%%%%%%%%%%%%%%%%%%%%%%%%%%%%%%%%%%%%%%%%%%%%%%%%%%%%%%%%%%%%%%%%%%%%%%%%%%%%%%%%%%%%%%%%%%%%%%%%%%%%%%%%%%%%%%%%%%%%
\begin{figure}
    \centering
    \includegraphics[width=\linewidth]{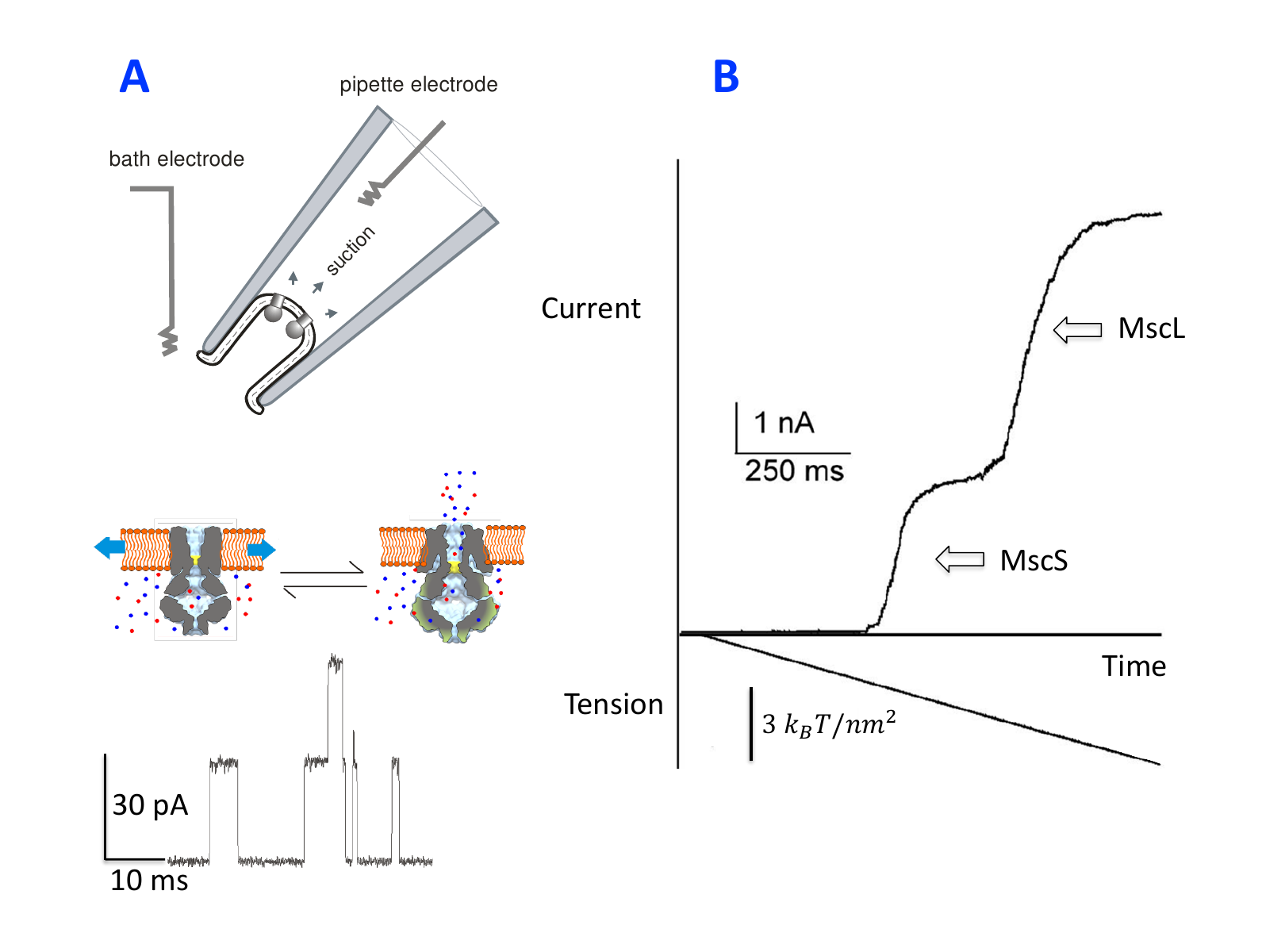}
    \caption{ \textbf{A} Cartoon of experimental set up for the excised patch configuration. Tightly sealed membrane provides an electrical isolation. However, activation of single channels in response to membrane tension forms conducting pathways which can be monitored with pico-ampere precision (depicted below). Observations of discrete currents passing through individual channels made patch-clamp one of the very first single-molecule technique. \textbf{B} Electrical activity of an excised patch of native \textit{E. coli} membrane in response to a linear increase in the membrane tension (adapted from \cite{ccetiner2017tension}). The two waves of current represent the activation of MscS channels at lower tension, followed by MscL channels at higher tension. If we zoom in on these current waves, we will see that they consist of discrete units, as shown in Panel A. Specifically, this patch contains around 60 MscS channels and 50 MscL channels. The scale bar shows current in nanoamperes (nA) or picoamperes (pA), time in unit of milliseconds (ms) and tension in units of $k_BT/nm^2$. Note that 1 $k_BT/nm^2=4.114\ mN/m$.} 
    \label{F2}
\end{figure}
\section*{Mechanosensation}
\subsection*{A brief history}
The history of gated pores dates back almost a century. In 1934, T. Kamada made the first intracellular electrical recordings from a living organism, using Paramecium \cite{kamada1934some}. A decade later, in 1945, Hodgkin and Huxley published their groundbreaking research on resting and action potentials in the squid giant axon \cite{hodgkin1945resting}, which led to their Nobel Prize in Physiology or Medicine in 1963. Building on this foundation, Erwin Neher and Bert Sakmann developed the patch clamp technique and revolutionized the ion channel research \cite{neher1976single}. Patch clamping involves clamping a piece of membrane with a polished glass micropipette to create a giga-ohm seal, providing electrical isolation across the membrane (see Fig. \ref{F2}A). This technique works simply because biological membranes are virtually impermeable to ions \cite{haydon1972ion,mueller1962reconstitution}. As a result, the exchange of ions between the inside of the pipette and the outside can only occur through the ion channels embedded in the membrane. When an ion channel opens, the movement of ions can be monitored as an electric current with picoampere precision, as illustrated in Fig. \ref{F2}A.

The idea of force-gated ion channels emerged in the late 1970s when Corey and Hudspeth reported rapid currents in bullfrog hair cells in response to hair bundle displacement \cite{corey1979response}. Subsequent research found tension-activated channels in chicken myoblasts \cite{guharay1984stretch}, yeast \cite{gustin1988mechanosensitive}, amphibians \cite{zhang2000mechanically}, and plants \cite{cosgrove1991stretch}. Bacterial cells, however, posed a challenge due to their small size and thick cell walls, making it difficult to study their cytoplasmic membranes with the patch-clamp technique. An important progress was made in 1985 when Ruthe and Adler grew \textit{E. coli} cells in the presence of cephalexin, which inhibits septum formation between cells and causes them to elongate into filaments \cite{ruthe1985fusion}. These filaments were treated with lysozyme to digest the cell walls, collapsing them into large spheroplasts between 2-10 $\mu m$ in diameter. Subsequently, in 1987, Martinac and colleagues recorded mechanosensitive ion channel activity in native \textit{E. coli} membranes using the patch-clamp technique \cite{martinac1987pressure}. The cytoplasmic membrane of \textit{E. coli} contains at least two distinct tension-activated channels, similar to what is shown in Fig. \ref{F2}B \cite{sukharev1993two}. As the membrane tension increases linearly with time, MscS channels open first. The second wave of activity is due to MscL channels opening, which require more tension to activate. In 1994, Sukharev isolated and cloned the \textit{mscL} gene through biochemical purification \cite{sukharev1994large}. Later, Levina and colleagues, using homology to the potassium efflux protein KefA, isolated the \textit{yggB} gene \cite{levina1999protection}, which encodes MscS.

In eukaryotes, mechanosensitive ion channels are crucial for various processes such as touch, sensing, and hearing. Examples include epithelial sodium channels (DEG/ENaC family), which are found in many animals that are involved in mechanotransduction in neurons \cite{arnadottir2010eukaryotic}, transient receptor potential (TRP) channels that help detect pain, temperature changes, pressure, and vision \cite{minke2002trp,ramsey2006introduction}, and the two-pore domain potassium channel family \cite{patel1998mammalian}. The recently discovered Piezo membrane proteins play key roles in processes like touch, breathing, and vascular development as well \cite{coste2012piezo}. For more details, readers can refer to several reviews on mechanosensation in higher organisms \cite{haswell2011mechanosensitive,arnadottir2010eukaryotic,sukharev2022mechanosensitive,kung2005possible}.
\subsection*{Gating mechanisms}
\begin{figure}
    \centering
    \includegraphics[width=\linewidth]{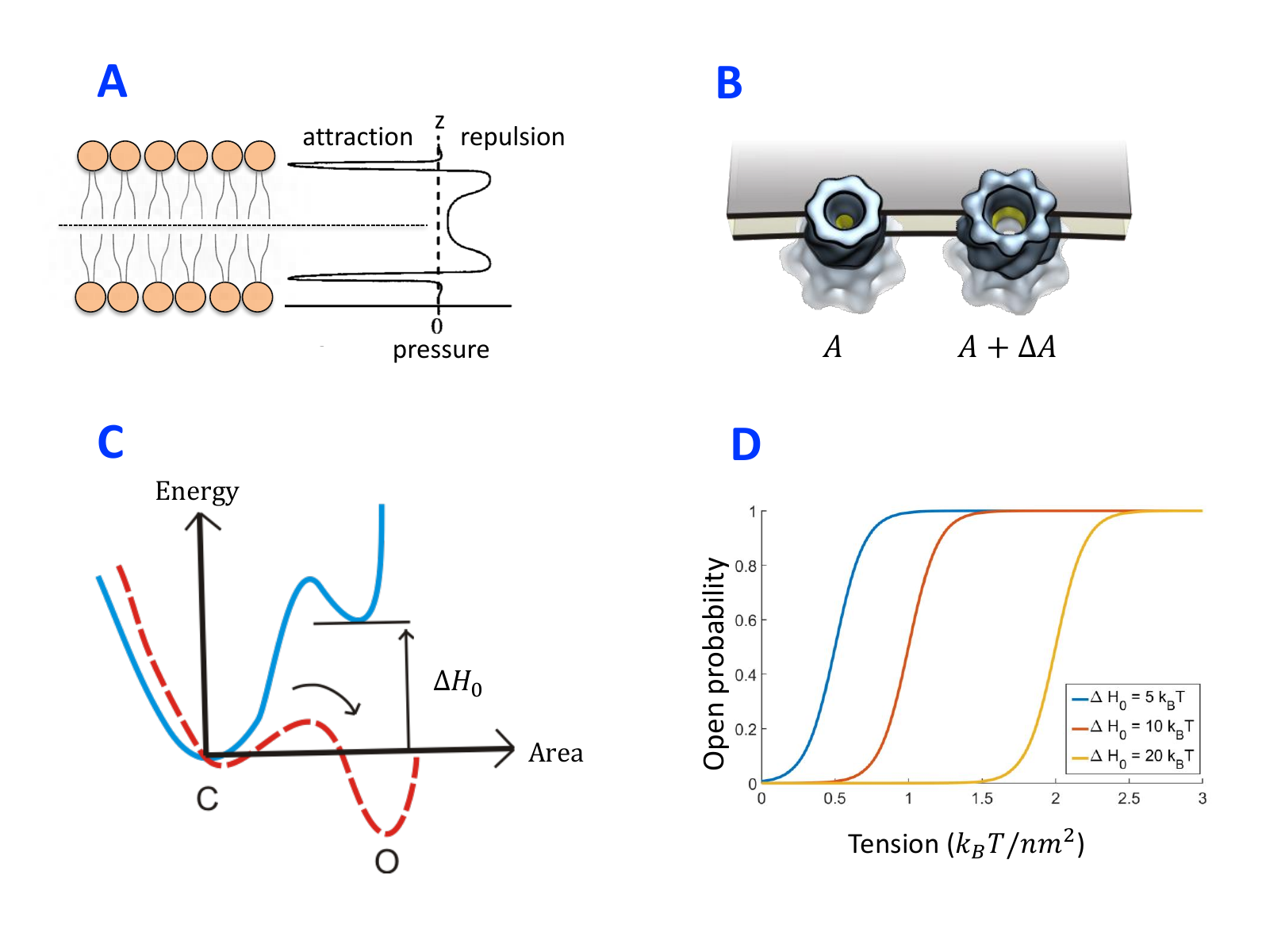}
    \caption{\textbf{A} Pressure profile of membranes. Membrane proteins experience highly anisotropic pull and push forces. Stretching the membrane can change the profile pressure further and lead to conformational changes, such as the opening of a mechanosensitive ion channel. \textbf{B} Force-from-lipid mechanism. The channel transitions from a closed state to an open state in response to an increase in membrane tension. No tether, cytoskeleton or any additional force-transmitting element is required. Open states have a larger in-plane expansion area (adapted from \cite{ccetiner2020recovery}). \textbf{C} Channels naturally reside in the closed state, which is the lower-energy state (blue curve). However, as the tension increases, the open state can become energetically more favorable (red dashed curve). \textbf{D} The equilibrium open probability of a two-state channel, Eq.\ref{popen2}, as a function $\Delta H_0$, for $\Delta H_0=5, 10, 20\ k_BT$ at a fixed $\Delta A =10\ nm^2$.}
    \label{F3}
\end{figure}
MS channels display a diverse range of structures and functions. Some require lots of membrane tension to open, while others require less. Some allow the passage of ions non-selectively by forming big pores, others favor negatively charged ions over positively charged ones. Despite all of these differences, which possibly originate from their evolution, there are striking similarities in what and how they do. 

 Life has found ways to manipulate the energy levels of ion channels using  mechanical forces such as membrane tension. But the underlying physics is far from trivial. Molecular dynamics simulations and thermodynamic models have been helpful in understanding the MS channels-lipid interactions \cite{cantor1997lateral,gullingsrud2004lipid,anishkin2010hydration,anishkin2014feeling}. A good starting point is the pressure profile in lipid membranes, which bear highly anisotropic forces, as shown in Fig. \ref{F3}A. Similar to surface tension at the interface of water and oil, the two polar-nonpolar interfaces at the level of the lipid ester bonds possess a large negative pressure. The strong repulsions among the hydrocarbon chains give rise to positive pressure, which decreases toward the midplane of the membrane. An additional contribution to this positive pressure comes from the repulsion between polar head-group interactions. Since the membrane is self-assembled and stable, these positive and negative forces balance each other \cite{cantor1997lateral}. Consider an MS channel that stays in a closed conformation inside the membrane. It already experiences anisotropic pull and push forces. Any additional stimulus that changes the membrane's pressure profile, such as applied tension \cite{anishkin2014feeling} or the insertion of an amphipath \cite{martinac1990mechanosensitive}, can lead to imbalances that favor the open state. Despite the complexity of interactions at the molecular level, we can still make some progress by reducing these interactions into a few experimentally measurable parameters to describe the gating of MS channels.  
 
 Let us model a MS channel as an effective two-state system that can be in a closed or open state. What we have in mind, in particular, is a mesoscopic description such that each of the two states (open or closed) consists of lots of microscopic degrees of freedom that cannot be experimentally distinguished. The transition from an all-atom, microscopic view to a mesoscopic one requires coarse-graining. Here, we assume that the closed and open states have energies $\epsilon_{\text{closed}}$ and $\epsilon_{\text{open}}$, and occupy membrane areas $A_{\text{closed}}$ and $A_{\text{open}}$, respectively, as depicted in Fig. \ref{F3}B. In our calculations, the differences between energies ( $\Delta H_0 \equiv \epsilon_{\text{open}}-\epsilon_{\text{closed}}$) and areas ($\Delta A = A_{\text{open}} - A_{\text{closed}}$) will be more important than the absolute values of these quantities. Just as pressure stabilizes states with smaller volumes, the open state becomes energetically more favorable when the membrane is under tension. We represent this situation by $\epsilon_{\text{open}}-\gamma \Delta A$, indicating that the energy of the open state is lowered in the presence of membrane tension by an amount $-\gamma \Delta A$, where $\gamma$ is the membrane tension (see Fig. \ref{F3}C). In patch-clamp experiments, the membrane tension $(\gamma)$ is related to the applied
pressure $(p)$ through the radius of curvature of the patch ($r$) according to the law of Laplace, $\gamma=pr/2$ \cite{haswell2011mechanosensitive,sukharev2022mechanosensitive,moe2005assessment}. It is important to note that MS channels sense the membrane tension, not the applied pressure \cite{moe2005assessment}. 

In a typical patch-clamp experiment, all channels are initially in a closed state and the conductance is negligible. Then, the membrane tension is increased linearly to a value of $\gamma_{\tau}$ in some time $\tau$. Usually, the probability of finding channels open is (almost) 1 when the tension reaches its final value of $\gamma_{\tau}$. This experimental protocol is called \textit{ramp}. If the tension is reduced back to its starting value from $\gamma_{\tau}$ with the same rate, this protocol is called \textit{triangular ramp}. When a triangular ramp is employed, we refer to the section where the tension is increasing as the ``forward" process, and the section where the tension is decreasing as the ``backward" process. As a side note, it is possible to employ more sophisticated experimental protocols but the ramps are the most commonly used.  

So far, we have discussed how an ion channel can transition from a closed configuration to an open in response to an increase in membrane tension. This mechanism is known as force-from-lipid. The first experimental evidence for this mechanism came when purified MscL and MscS were reconstituted in artificial bilayers gated in response to applied tension \cite{sukharev1997mechanosensitive,nomura2012differential}. No tether, cytoskeleton or any additional force-transmitting element is needed. Some MS channels might also sense force from a tether. In this case, a mechanical stimulus is transmitted to the ion channel through a tether. This force can originate from fibrillar elements like the cytoskeleton or the extracellular matrix. Similar to how $-\gamma \Delta A$ term biases the open state, the presence of a force $f$ can favor energetically the open state by an additional work term, $-f\Delta l$, where $\Delta l$ is the tether displacement against the force $f$ \cite{anishkin2014feeling,sukharev2022mechanosensitive}. The conjugate pairs of variables, $\left(f, \Delta l \right)$ or $\left(\gamma, \Delta A \right)$ give us one or two dimensional work, and introduce a bias in favor of the open state. We should note that force-from-lipid or force-from-tether models are not mutually exclusive and growing experimental evidence shows that some channels might combine different mechanisms to gate. For example, eukaryotic two-pore-domain potassium channels such as TRAAK and TREEK1, when reconstituted into liposomes, are responsive to mechanical forces similar to MscL and MscS \cite{berrier2013purified,brohawn2014mechanosensitivity}. What is even more surprising is that a canonical voltage-dependent potassium channel channel, Kv, is also highly sensitive to membrane tension \cite{schmidt2012mechanistic}. All these suggest that the force-from-lipid paradigm is ancient and mostly preserved by eukaryotic MS channels \cite{anishkin2014feeling} and that MS channels can integrate information from various signals.
%%%%%%%%%%%%%%%%%%%%%%%%%%%%%%%%%%%%%%%%%%%%%%%%%%%%%%%%%%%%%%%%%%%%%%%%%%%%%%%%%%%%%%%%%%%%%%%%%%%%%%%%%%%%%%%%%%%%%%%%%%%%%%%%%%%%%%%%%%%%%%%%%%%%%%%%%%%%%%%%%%%%%%%%%%%%%%%%%%%%%%%%%%%%%%%%%%%%%%%%%%%%%%%%%%%%%%%%%%%%%%%%%%%%%%%%%%%%%%%%%%%%%%%%%%%%%%%%%%%%%%%%%%%%%%%%%%%%%%%%%%%%%%%%%%%%%%
\subsection*{An example of information integration}
\begin{figure}
    \centering
    \includegraphics[width=\linewidth]{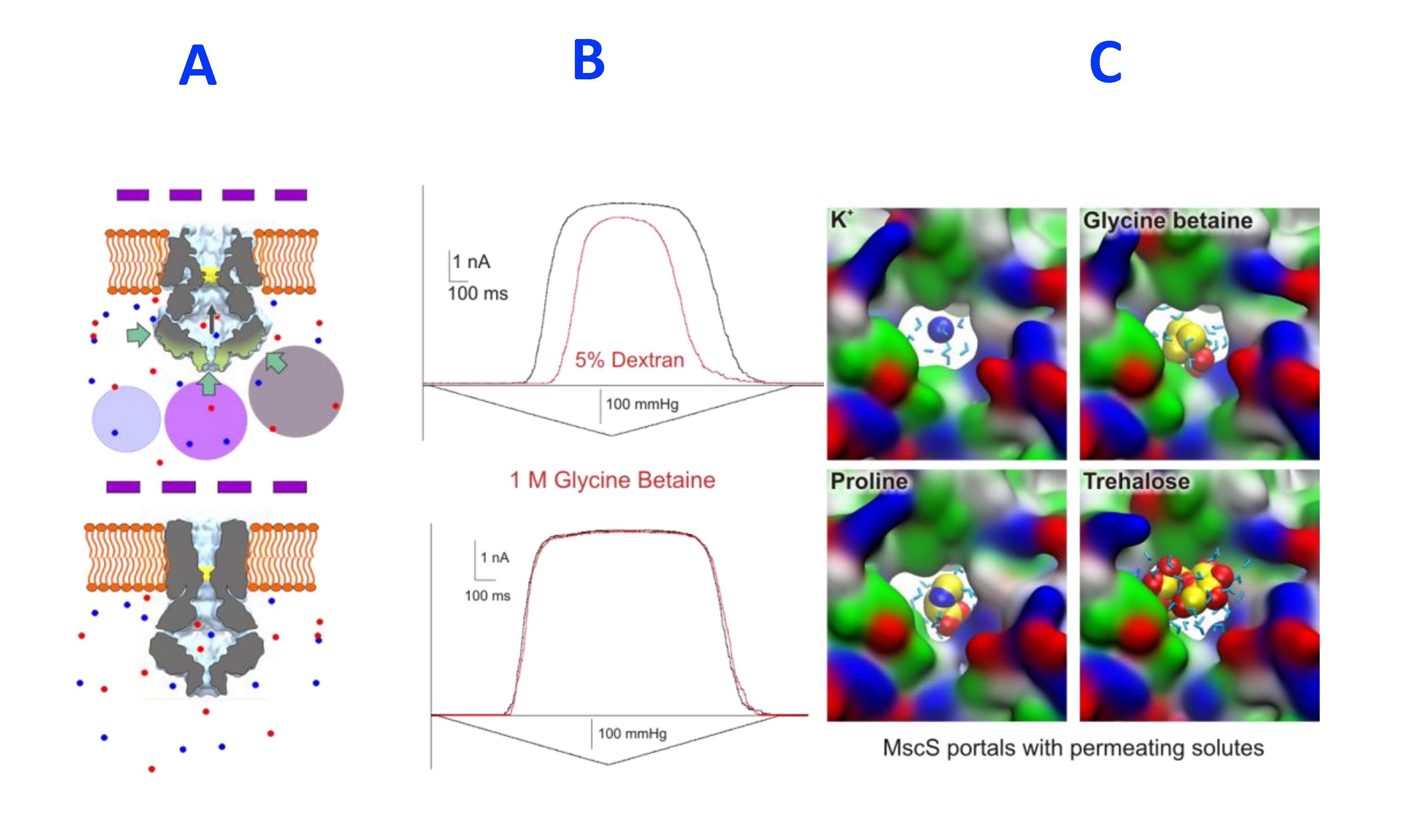}
    \caption{Size-dependent effect of crowding agents on the gating properties of MscS. \textbf{A} A cartoon shows that in the presence of big molecules that cannot penetrate through the cytoplasmic cage region of the channel, the difference between the chemical potentials of water molecules inside and outside of the cage creates an osmotic pressure ($\Pi_{\text{osm}}$) acting on the cytoplasmic domain of the channel. \textbf{B} Since the opening of a channel is also accompanied by the expansion of the cytoplasmic cage domain, this expansion, in the presence of cage-inaccessible molecules such as Dextran, comes with an extra energy penalty of the form $\Pi_{\text{osm}}\Delta V$. This extra work term favors the closed state as it excludes less volume to crowding agent and results in a right shift in the open probability of the channels. For ions and relative small solutes that are free to pass through the cage portals and equilibrate such as Glycine betaine, there is no net osmotic pressure action on the cage, thus the transition rates stay the same. \textbf{C} Molecular dynamics simulations reveal that small molecules such as ions, Glycine betaine and Proline can pass through the MscS portals, whereas Trehalose is expected to partially permeate through the cage. Therefore, the cage region would be inaccessible to the molecules that are bigger than Trehalose. These figures are adapted from \cite{cetiner2018nanoscale}.}
    \label{F4}
\end{figure}
Most channels may primarily respond to one type of stimulus, but their energy landscape can be influenced by different types of perturbations. Hence, there is lots of room for information integration. A good example is MscS, which can integrate information about cytoplasmic crowding into its gating behavior as follows. The cytoplasmic cage domain of MscS (see Fig. \ref{F4}A) expands when the channel opens, increasing its volume (denoted $\Delta V$) \cite{machiyama2009structural}. Keeping this in mind, in the presence of large macromolecules that cannot penetrate MscS’s cage, there is a difference in the chemical potential of the water molecules inside and outside the cage.
This creates an osmotic pressure, $\Pi_{\text{osm}}$, acting on the cage domain. Hence, this time, there is an energetic penalty of the form $\Pi_{\text{osm}} \Delta V$ when the channel tries to open. This cost arises from the work done against the osmotic pressure and disfavors the open state. One possible use of integrating cytoplasmic crowding information is that MscS senses the dehydration and prefers to shoot itself off loss to prevent further loss of water and internal osmolytes. 

This idea was tested by introducing large Dextran molecules on the cytoplasmic side of the channels and applying a triangular ramp protocol to see how the channel gating is affected. As demonstrated in Fig. \ref{F4}B, in the presence of cage-inaccessible (large) Dextran molecules, the channels required more tension to gate during the forward process (red curve, ascending leg). Additionally, the channels tended to close more readily during the backward process (red curve, descending leg). These results are consistent with the previous studies \cite{rowe2014cytoplasmic,grajkowski2005surface}. On the other hand, small molecules like Glycine betaine, which can freely move through the cage, did not alter the channels' gating behavior because they cannot create a significant difference in water potential across the cage. As a result, the response to a triangular ramp remained unchanged (Fig. \ref{F4}B). In accordance with the molecular dynamic simulations that probed the permeability of various crowding agents into the cytoplasmic cage of MscS channels (Fig. \ref{F4}C), we found that for the molecules that can freely diffuse through the cage, there is no significant change in the transition rates of the channel \cite{cetiner2018nanoscale}. The full thermodynamic analysis will be published elsewhere. 
%%%%%%%%%%%%%%%%%%%%%%%%%%%%%%%%%%%%%%%%%%%%%%%%%%%%%%%%%%%%%%%%%%%%%%%%%%%%%%%%%%%%%%%%%%%%%%%%%%%%%%%%%%%%%%%%%%%%%%%%%%%%%%%%%%%%%%%%%%%%%%%%%%%%%%%%%%%%%%%%%%%%%%%%%%%%%%%%%%%%%%%%%%%%%%%%%%%%%%%%%%%%%%%%%%%%%%%%%%%%%%%%%%%%%%%%%%%%%%%%%%%%%%%%%%%%%%%%%%%%%%%%%%%%%%%%%%%%%%%%%%%%%%%%%%%%%%%%%%%%%%%%%%%%%%%%%%%%%%%%%%%%%%%%%%%%%%%%%%%%%%%%%%%%%%%%%%%%%%%%%%%%%%
\subsection*{Thermodynamic equilibrium}
Equilibrium formalism is unmatched in its simplicity and applicability. When a system reaches thermodynamic equilibrium, all the specifics of the underlying microscopic dynamics become irrelevant. If we know the state energies, we can calculate the probability of observing a given state, from which all relevant statistical properties can be obtained. For example, when the membrane tension is held fixed at $\gamma$, the equilibrium probability of finding a channel in the open state is given by the following Boltzmann distribution, 
\begin{equation} \label{popen}
 \text{Pr}_{\text{open}}(\gamma)=\frac{\mathrm{e}^{-\beta \left(\epsilon_{\text{open}}-\gamma \Delta A \right)}}{Z(\gamma)},
\end{equation}
where $\beta=1/(k_BT)$. The partition function, $Z(\gamma)$, is parameterized by the membrane tension and is defined as follows, $Z(\gamma)=\mathrm{e}^{-\beta \left(\epsilon_{\text{open}}-\gamma \Delta A \right)}+\mathrm{e}^{-\beta \epsilon_{\text{closed}}}$ for a two-state channel. In a typical patch-clamp experiment, the temperature of the buffer solution may be different from the room temperature since these buffer solutions are usually kept in cold fridges. Hence, care must be taken when running experiments. The safest option is to wait until the buffer solution reaches the room temperature to prevent side effects due to temperature differences. 

It is often more convenient to write the Eq.\ref{popen} as follows,
\begin{equation}\label{popen2}
    \text{Pr}_{\text{open}}(\gamma)=\frac{1}{1+\mathrm{e}^{\beta (\Delta H_0-\gamma \Delta A)}}.
\end{equation}
Fig. \ref{F3}D, displays the equilibrium probability of the channel being in the open state as a function of membrane tension for three different $\Delta H_0=5,10,20\ k_BT$ at a fixed $\Delta A=10\ nm^2$. As $\Delta H_0$ increases, the channel requires more tension to open, which shifts the open probability curve to the right. This curve, showing the open probability as a function of the stimulus, is one of the key pieces of data when studying ion channels. It is known as a ``dose-response curve.” It is often the case that Eq.\ref{popen2} is fitted to an experimental dose-response curve to extract the values of $\Delta H_0$ and $\Delta A$. But there is a catch---this approach assumes the system is at thermodynamic equilibrium. Unless the ramp is applied very slowly, making it almost like a quasi-static process, this assumption does not hold. Indeed, during a ramp protocol, where tension is increased linearly, the channels may be driven out of equilibrium, which we discuss next.   
%%%%%%%%%%%%%%%%%%%%%%%%%%%%%%%%%%%%%%%%%%%%%%%%%%%%%%%%%%%%%%%%%%%%%%%%%%%%%%%%%%%%%%%%%%%%%%%%%%%%%%%%%%%%%%%%%%%%%%%%%%%%%%%%%%%%%%%%%%%%%%%%%%%%%%%%%%%%%%%%%%%%%%%%%%%%%%%%%%%%%%%%%%%%%%%%%%%%%%%%%%%%%%%%%%%%%%%%%%%%%%%%%%%%%%%%%%%%%%%%%%%%%%%%%%%%%%%%%%%%%%%%%%%%%%%%%%%%%%%%%%%%%%%%%%%%%%%%%%%%%%%%%%%%%%%%%%%%%%%%%%%%%%%%%%%%%%%%%%%%%%%%%%%%%%%%%%%%%%%%%%%%%%
\section*{Hysteresis: A signature for non-equilibrium behaviour}
\begin{figure}
    \centering
    \includegraphics[width=\linewidth]{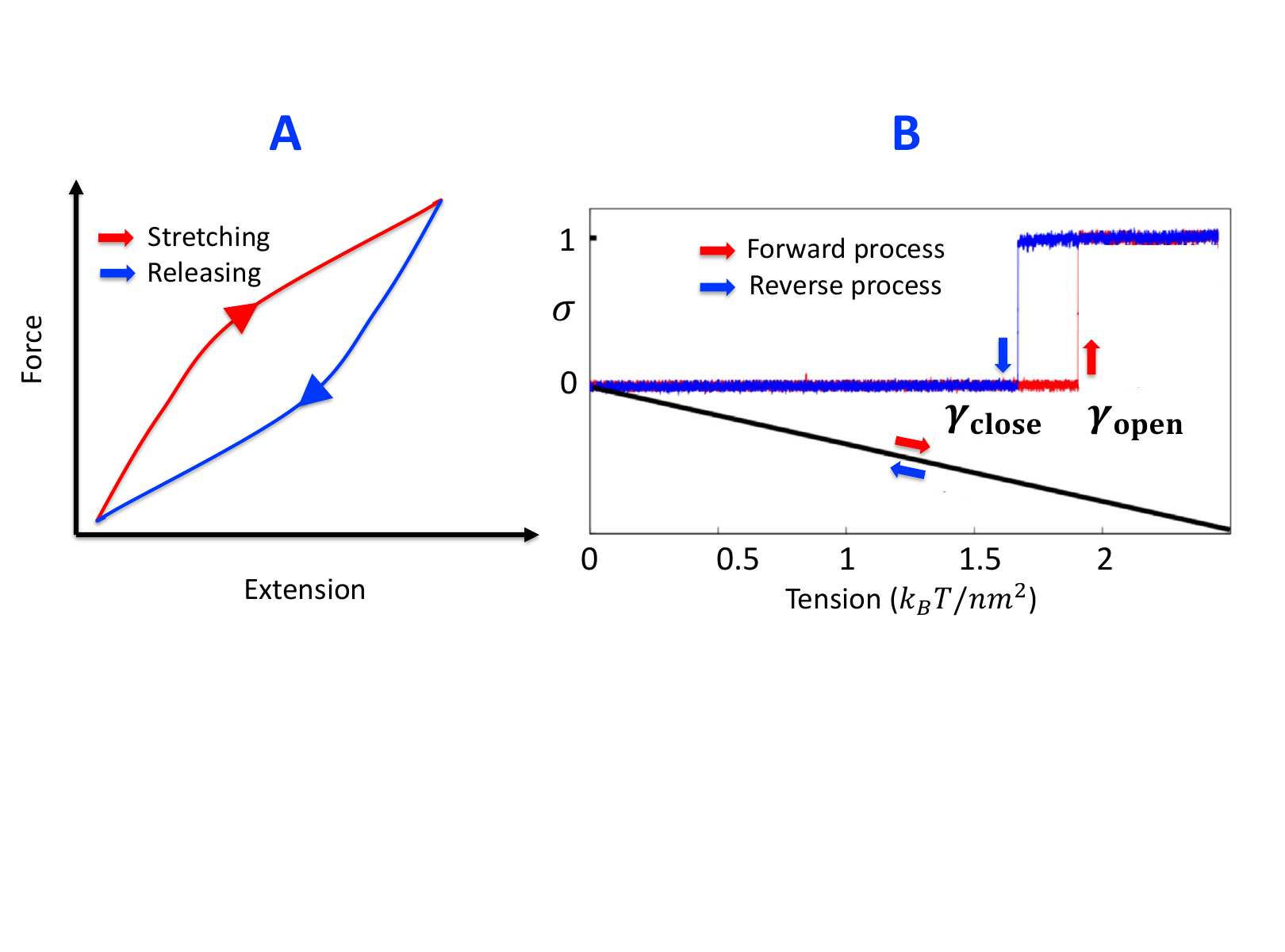}
    \vspace*{-40mm}
    \caption{\textbf{A} A rubber band is an example of a driven macroscopic system. When a rubber band is stretched (red curve) and then released (blue curve) at the same rate, it does not follow the same path in the force-extension curve, giving rise to a hysteresis. The area between these curves is a measure of energy lost as heat into the environment. \textbf{B} Nanoscale hysteresis is observed when a MscS channel is driven by membrane tension in a cyclic manner (adapted from \cite{ccetiner2020recovery}). Here, $\sigma$ is the state variable, which is 1 when the channel is in the open state and 0 when the channel is closed. As the tension is increased linearly, the channel opens at $\gamma_{\text{open}}$ (indicated by the red arrow). We call this process forward (red trace). In the reverse process, the tension is reduced back to its initial value with the same rate (blue trace), and the channel closes at $\gamma_{\text{close}}$. The area between these two curves is again proportional to the heat dissipated into the environment during this thermodynamic process.}
    \label{F5}
\end{figure}
The laws of thermodynamics, when applied to small systems like an MS channel, yield unexpected and surprising results compared to macroscopic systems like a rubber band. In this section, our analysis is more phenomenological than quantitative.

Consider a thermodynamic process during which a system is driven from one equilibrium state ($A$) to another ($B$) by changing a control parameter according to a protocol. In the case of a rubber band, the control parameter may be the length of it. By changing its length, we can do work on the rubber band. First, let us hold the rubber band at length $A$ and wait for it to reach equilibrium at room temperature, say $T$. Then, we stretch the rubber band to a new length $B$ and wait for it to reach equilibrium again at this length. Let us arbitrarily call this process \emph{forward}. The work done during the forward process, denoted as $W^F$, is always greater than or equal to the Helmholtz free energy differences $\Delta F$:
\begin{equation}\label{Secondlaw}
W^F\geq \Delta F= F_{\text{B,T}}-F_{\text{A,T}},
\end{equation}
where $F_{\text{X,T}}$ is the Helmholtz free energy of the rubber band, corresponding to the length $X$ and temperature $T$.
If the rubber band is pulled slowly enough to follow a sequence of equilibrium states during its evolution, then Eq.\ref{Secondlaw} becomes an equality. Let us also consider the reverse process, where the rubber band, starting from an equilibrium state B, is released back to length $A$ with the same rate. The work performed in this process is denoted by $W^R$ and is subject to similar bounds given in Eq.\ref{Secondlaw}, $W^R\geq F_{\text{A,T}}-F_{\text{B,T}}$. 
For a cyclic process during which the system starts and ends in the same equilibrium state, we see that
\begin{equation}\label{Cyclic}
W^F+W^R \geq 0,
\end{equation} 
which means there is no \textit{free lunch} in nature. This effect manifests itself as a hysteresis in the force-extension curve of a rubber band. The rubber band heats up when it is stretched, and it requires more force to pull, accordingly, the work done during stretching is always greater than the release. The area between the stretching and release curves in the force-extension diagram is the heat dissipated into the environment, see Fig. \ref{F5}A. Therefore, hysteresis is a signature of non-equilibrium behavior in driven systems. 

We can observe hysteresis at the nanoscale as well. Just as we can stretch a rubber band according to a protocol, we can also stretch ion channels using the patch-clamp technique. Consider the traces shown in Fig. \ref{F5}B, where the membrane tension is increased linearly during the forward process and then decreased back to the initial value at the same rate during the reverse process. These traces, which represent the gating of MscS channels, clearly show hysteresis. So, MscS channels can be driven away from equilibrium during a typical experimental protocol.
A natural follow-up question is: how much heat is dissipated into the environment during these processes? By quantifying the dissipated heat, we can understand the energetic efficiency of ion channels and how underlying parameters, such as $\Delta H_0$ and $\Delta A$, have been shaped by billions of years of evolution. But first, we need to define the stochastic thermodynamics of ion channels.
%%%%%%%%%%%%%%%%%%%%%%%%%%%%%%%%%%%%%%%%%%%%%%%%%%%%%%%%%%%%%%%%%%%%%%%%%%%%%%%%%%%%%%%%%%%%%%%%%%%%%%%%%%%%%%%%%%%%%%%%%%%%%%%%%%%%%%%%%%%%%%%%%%%%%%%%%%%%%%%%%%%%%%%%%%%%%%%%%%%%%%%%%%%%%%%%%%%%%%%%%%%%%%%%%%%%%%%%%%%%%%%%%%%%%%%%%%%%%%%%%%%%%%%%%%%%%%%%%%%%%%%%%%%%%%%%%%%%%%%%%%%%%%%%%%%%%%%%%%%%%%%%%%%%%%%%%%%%%%%%%%%%%%%%%%%%%%%%%%%%%%%%%%%%%%%%%%%%%%%%%%%%%%
\section*{Stochastic thermodynamics of ion channels}
 Every patch-clamp recording is a stochastic trajectory. The recent advancements in the non-equilibrium physics of small systems now enable us to calculate how much heat is observed from the surrounding environment, denoted by $Q$, how much work is done on the system, denoted by $W$ at the level of as single stochastic trajectory. And, of course, they are related to each other through energy conservation:
\begin{eqnarray}
    \Delta H= W+Q,
\end{eqnarray}
where $\Delta H$ is the change in the energy of the system. To define work and heat for ion channels, it will be convenient to introduce a state variable, $\sigma$, which is 0 for a closed channel and 1 for an open one. Let $\epsilon_{\text{closed}}$ and $\epsilon_{\text{open}}$ denote the energies of the closed and open states, as before. The energy of the system, in the presence of tension $\gamma$, can be written in a compact form as follows,
\begin{equation}\label{Energy}
H(\sigma,\gamma)=H_0(\sigma)-\gamma A(\sigma),
\end{equation}
where $H_0(\sigma)=(1-\sigma)\epsilon_{\text{closed}}+\sigma\epsilon_{\text{open}}$ represents the energy of the system due to the state of the ion channel alone, and the additional term $\gamma A(\sigma)=\gamma \sigma \Delta A$
represents the energy difference between open and closed channels in the presence of tension, where $\Delta A$ is the area difference between the closed and open states of the channel. 

Let us consider a typical ramp protocol, where the membrane tension is increased from an initial value of zero to a final value of $\gamma_{\tau}$ during time $\tau$. Let us call this process forward as usual. For a single realization of an experimental protocol of duration $\tau$, the work performed on the channel is given by
\begin{equation}\label{work}
W\equiv \int_0^{\tau} \dot{\gamma} \frac{\partial H}{\partial \gamma} dt,
%= -\Delta A \int_0^{\tau} \dot{\gamma} \sigma dt,
\end{equation}
where $\dot{\gamma}=d\gamma/dt$ and the heat absorbed from the environment is given by 
\begin{equation}\label{heat}
Q\equiv \int_0^{\tau} \dot{\sigma} \frac{\partial H}{\partial \sigma} dt.
%= \int_0^{\tau} (\epsilon_{\text{open}}-\epsilon_{\text{closed}}-\gamma \Delta A) \dot{\sigma} dt. 
\end{equation}
In stochastic thermodynamics, work done on a system is defined as the change in the internal energy resulting from the manipulation of the control parameter (tension), which is captured by Eq.\ref{work}. Whereas, heat is defined as the energy exchanged between the system and its thermal environment (reservoir) due to stochastic jumps in the system's states, which is given by Eq.\ref{heat}. It may be instructive to calculate these quantities for the trajectory obtain during the forward process, as shown in Fig. \ref{F5}B. In this trajectory, a single MscS channel transitioned from closed to open state at $\gamma_{\text{open}}$ in response to a ramp protocol. Since this transition is quite fast, the state of the channel, represented by $\sigma$, can be approximated with a Heaviside step function, $\sigma=\theta(\gamma-\gamma_{\text{open}})$. Namely:
\begin{equation}\label{state}
 \sigma=
\begin{cases}
1, & \gamma \geq \gamma_{\text{open}} \\
0, & \gamma < \gamma_{\text{open}}.  
\end{cases} 
\end{equation}
The step function approximation of the state of the channel allows us to calculate heat and work explicitly. Let us start with the work done on the system,
\begin{equation}\label{workdetailed}
\begin{split}
W\equiv \int_0^{\tau} \dot{\gamma} \frac{\partial H}{\partial \gamma} dt &= -\Delta A \int_0^{\tau} \dot{\gamma} \sigma dt =-\Delta A \int_0^{\gamma_{\tau}} d\gamma \sigma \\
&=-\Delta A \int_0^{\gamma_{\tau}} d\gamma \theta(\gamma-\gamma_{\text{open}})=-\Delta A (\gamma_{\tau}-\gamma_{\text{open}})
\end{split}
\end{equation}
In moving from the first line to second, we replaced $\sigma$ with the Heaviside step function (Eq.\ref{state}) and used the properties of step function to get the last equality. Eq.\ref{workdetailed} implies that the work done is proportional to the area under red curve in Fig. \ref{F5}B.  Similarly, using the relation between step function and the Dirac-Delta function, $d\sigma=\delta(\gamma-\gamma_{\text{open}})d\gamma$, we can express heat received from the environment as follows,
\begin{equation}\label{heatdetailed}
\begin{split}
Q&\equiv \int_0^{\tau} \dot{\sigma} \frac{\partial H}{\partial \sigma} dt =\int_0^{\gamma_{\tau}}\frac{\partial H}{\partial \sigma} \delta(\gamma-\gamma_{\text{open}}) d\gamma\\
& =\int_0^{\gamma_{\tau}} (\epsilon_{\text{open}}-\epsilon_{\text{closed}}-\gamma \Delta A)\delta(\gamma-\gamma_{\text{open}})d\gamma= \epsilon_{\text{open}}-\epsilon_{\text{closed}}-\gamma_{\text{open}}\Delta A
\end{split}
\end{equation}
We have calculated the work done the channel and heat received from the environment during the forward process for the patch-clamp trace in Fig. \ref{F5}B. For completeness, let us also obtain the change in the energy of the system. In the beginning of the protocol, the channel is in the closed state ($\sigma=0$) and tension is zero $\gamma=0$. At the end of the protocol, the channel is in the open state ($\sigma=1$) and the tension is $\gamma_\tau$. Therefore, using the definition given in Eq.\ref{Energy}, the total change in the energy is:
\begin{equation}\label{deltah}
\Delta H=H(\sigma=1,\gamma=\gamma_{\tau})-H(\sigma=0,\gamma=0)
=\epsilon_{\text{open}}-\epsilon_{\text{closed}}-\gamma_{\tau}\Delta A.
\end{equation}
Summing up Eqs. \ref{workdetailed} and \ref{heatdetailed}, we can show that $\Delta H=W+Q$, as expected.
\begin{equation}\label{firstlaw}
\begin{split}
    W+Q= & [-\Delta A (\gamma_{\tau}-\gamma_{\text{open}})]+[\epsilon_{\text{open}}-\epsilon_{\text{closed}}-\gamma_{\text{open}}\Delta A] \\
    &=\epsilon_{\text{open}}-\epsilon_{\text{closed}}-\gamma_{\tau}\Delta A=\Delta H
\end{split}
\end{equation}
Just in the case of a rubber band, we can also show that in a cyclic process, the total work done is non-negative. The reverse process, where decreased back to zero with the same rate, is given by the blue trace in Fig. \ref{F5}B. The channel that opened at $\gamma_{\text{open}}$ during the forward process and closes at $\gamma_{\text{close}}$ during the reverse one. The total work done during the forward process, as we calculated previously, is $W^F=-\Delta A (\gamma_\tau-\gamma_{\text{open}})$ and similarly, the work performed on the reverse process, is $W^R=\Delta A(\gamma_\tau-\gamma_{\text{close}})$. Hence, he total work during the cyclic process, 
\begin{equation}
    W^F+W^R=-\Delta A (\gamma_\tau-\gamma_{\text{open}})+\Delta A(\gamma_\tau-\gamma_{\text{close}})=\Delta A (\gamma_{\text{open}}-\gamma_{\text{close}}) >0
\end{equation}
Note that the trajectory in Fig. \ref{F5}B is a typical trajectory--this is what you would see mostly likely if you were the repeat the same experimental protocol. However, at nanoscale, thermal fluctuations can give rise to rare, second-law violating trajectories. Consider, for the moment, the colors are reversed in Fig. \ref{F5}B and what we called is forward previously now becomes reverse and vice versa. In this trajectory, the total dissipation, measured by $W^F+W^R$, would be negative. In such a scenario, the random jiggling inside the surrounding reservoir work in favor of the channel, at it requires less work to open by borrowing some energy in the form of heat from the surrounding reservoir. These rare trajectories are inevitable at nanoscale. However, there is still no free lunch but this time on average, 
\begin{equation}\label{workcyclic}
    \langle W^F\rangle +\langle W^R\rangle \geq 0,
\end{equation}
where the angular brackets represent an average over the corresponding distributions of work values obtained during the forward ($\Pr(W^F)$) and reverse processes ($\Pr(W^R)$), e.g., $\langle W^F\rangle \equiv \int W^F \Pr(W^F) \mathrm{d} W^F$. We can consider Eq.\ref{workcyclic} as the nanoscale analog of Eq.\ref{Cyclic}. 
\begin{figure}
    \centering
    \includegraphics[width=\linewidth]{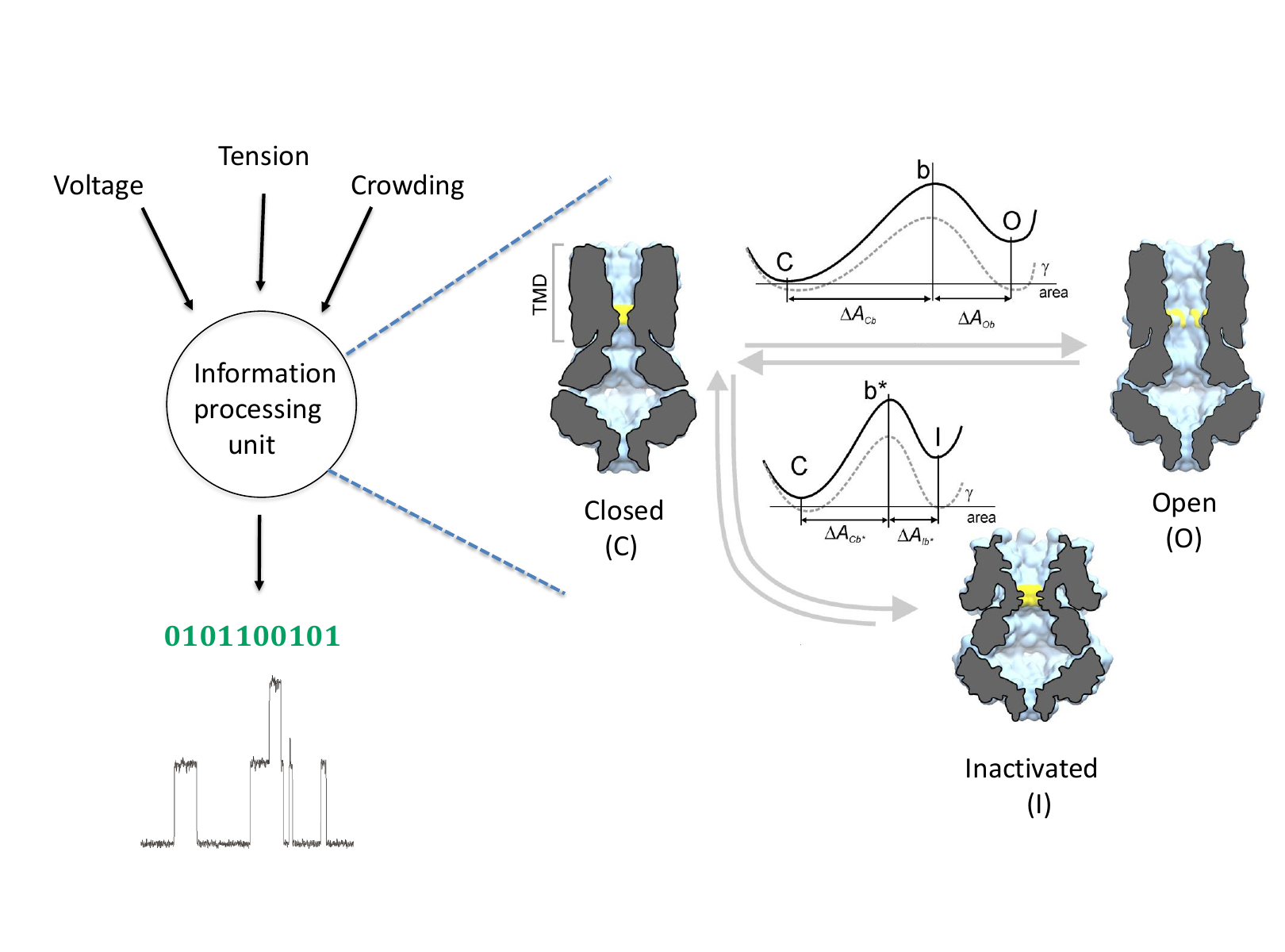}
    \caption{Mechanosensitive ion channels can integrate information from multiple inputs, such as tension and molecular crowding, process this information using their own dynamics, and finally output some information in the form of ionic currents, which we represent simply as 0 for a closed state and 1 for an open state. Downstream events may couple with these ionic currents, thereby the propagation of information may continue. On the right, we zoom in on the information processing unit's core, which is the energetic landscape of MscS. The discrete conformational state space of allowed transitions is indicated by the experiments. An Arrhenius type of relation describes the transition rate from state $y$ to state $x$: $k_{xy}=k^0_{xy}\exp(\beta \gamma \Delta A_{yb})$, where $k^0_{xy}$ is the intrinsic rate of the system's attempts to jump over the barrier between $x$ and $y$ in the absence of tension, and $\Delta A_{yb}$ is the expansion area from state $y$ to the barrier. The barriers that separate the open and inactivated states from the closed state are denoted by $b$ and $b^*$, respectively. The following parameters were used for simulating the MscS, $k^0_{CO}=9897\ s^{-1}$, $k^0_{OC}=4\mathrm{e}-6\ s^{-1}$, $k^0_{IC}=2\mathrm{e}-5\ s^{-1}$, $k^0_{CI}=0.18\ s^{-1}$, $|\Delta A_{Cb}|=7\ nm^2$, $|\Delta A_{Ob}|=5\ nm^2$, $|\Delta A_{Cb^*}|=5\ nm^2$, and $|\Delta A_{Ib^*}|=1.2\ nm^2$.}
    \label{F6}
\end{figure}
\subsection*{Fluctuation theorems and Landauer's principle}
We may think that the fluctuations around averages in a small system are just random noise coming from the system's interactions with the surrounding thermal reservoir. But this is not the whole story. These fluctuations encode very useful information and obey very strong symmetry relations. In 1997, Jarzynski showed that it is possible to recover the equilibrium free energy difference $(\Delta F)$ from non-equilibrium work distributions using the formula below \cite{jarzynski1997nonequilibrium},
\begin{equation}
    \mathrm{e}^{-\beta \Delta F}= \langle \mathrm{e}^{-\beta W} \rangle.
\end{equation}
There is more. The distributions of the work values obtained during the forward and reverse processes satisfy the following symmetry relation, which is known as Crooks fluctuation theorem \cite{crooks1999entropy}, 
\begin{equation}
    \frac{\Pr(W^F)}{\Pr(-W^R)}=\mathrm{e}^{\beta(W^F-\Delta F )}.
\end{equation}
Finally, there is the Landauer principle, which states that erasing a bit of information costs at least $k_BT \ln(2)$ in heat dissipation. Recently, the work and heat distributions have been obtained for MscS \cite{ccetiner2020recovery,ccetiner2023dissipation}. The work distribution of MscS has been used to obtain the free energy difference. In addition, the heat dissipated into the environment during the erasure of one bit of information encoded by the state of the channel ($\sigma=0$ or $\sigma=1$), has been shown to reach the Landauer bound for certain erasing protocols. What is interesting is that these protocols that reach Landauer's bound operate on timescales and force magnitudes similar to the natural conditions under which MscS evolved to function. In the next section, we go into details of how evolution has shaped the energy landscape of MscS.  
%%%%%%%%%%%%%%%%%%%%%%%%%%%%%%%%%%%%%%%%%%%%%%%%%%%%%%%%%%%%%%%%%%%%%%%%%%%%%%%%%%%%%%%%%%%%%%%%%%%%%%%%%%%%%%%%%%%%%%%%%%%%%%%%%%%%%%%%%%%%%%%%%%%%%%%%%%%%%%%%%%%%%%%%%%%%%%%%%%%%%%%%%%%%%%%%%%%%%%%%%%%%%%%%%%%%%%%%%%%%%%%%%%%%%%%%%%%%%%%%%%%%%%%%%%%%%%%%%%%%%%%%%%%%%%%%%%%%%%%%%%%%%%%%%%%%%%%%%%%%%%%%%%%%%%%%%%%%%%%%%%%%%%%%%%%%%%%%%%%%%%%%%%%%%%%%%%%%%%%%%%%%%%
\section*{Back to evolution and osmotic fitness}
\begin{figure}[h!!]
    \centering
    \includegraphics[width=\linewidth]{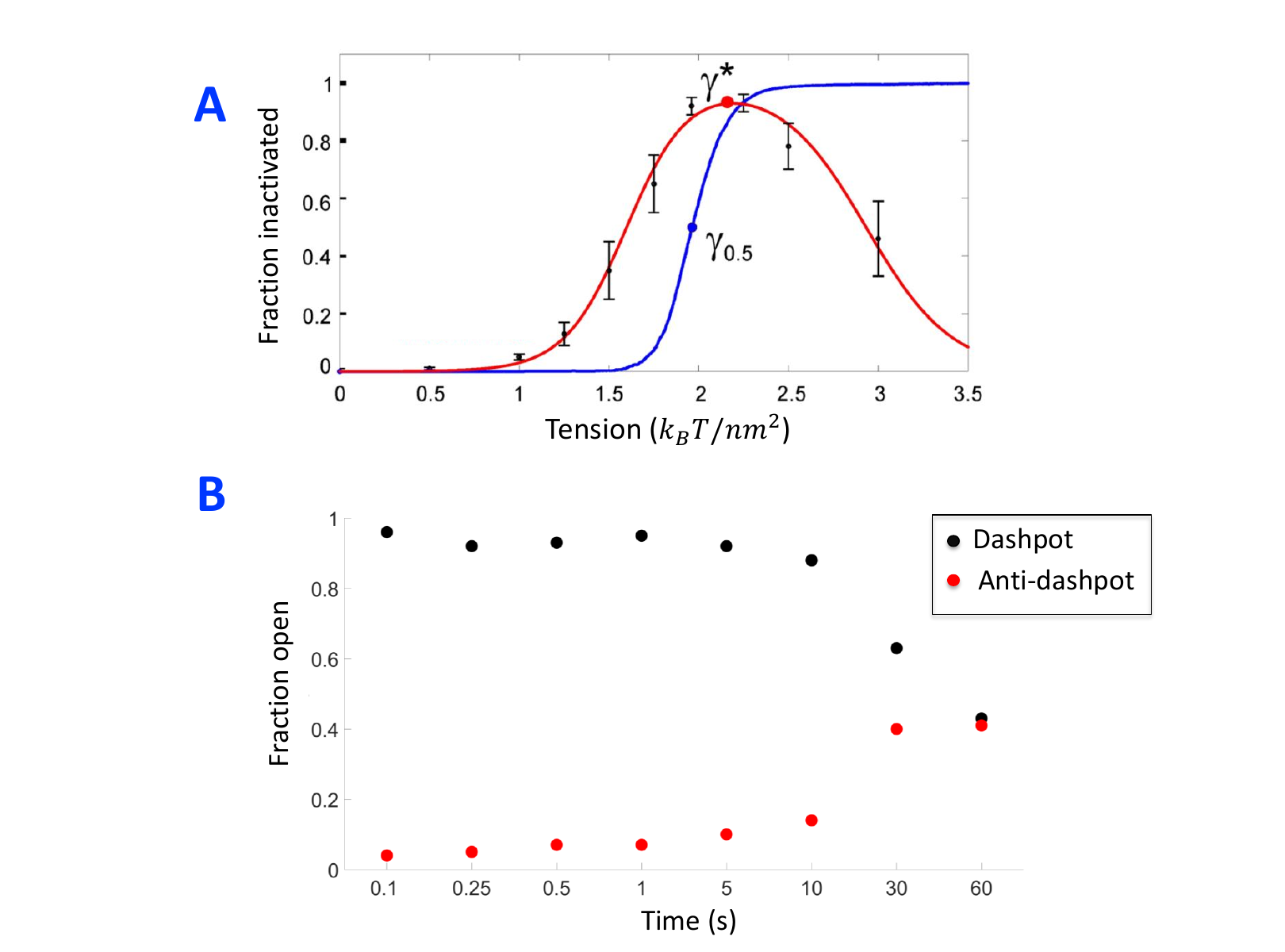}
    \caption{\textbf{A} Inactivation of MscS as a function of the membrane tension. All the data is generated using QUBexpress software available at \url{https://qub.mandelics.com}. The black data points with the error bars represent steady-state MscS inactivation during a 120-s constant tension, averaged over 8 independent patches. The red curve represents the amount of inactivation predicted by the three-state model. The red circle is the tension at which the steady-state inactivation is maximized. The blue curve is the open probability for the fast activation of channels under 1 s ramp protocol on which the midpoint, $\gamma_{0.5}$, is marked as blue circle. MscS does not inactivate when the tension is too low or too high. The sweet-spot is around the midpoint tension  (adapted from \cite{ccetiner2018spatiotemporal}). \textbf{B} The rate dependency of MscS inactivation. In the simulations, the membrane tension is increased from 0 to $3\ k_BT/nm^2$ at different rates, fastest in 0.1 second, slowest in 60 seconds. MscS channels tend to inactivate and ignore the slowly applied tension, as shown by the decrease in the number of open channels (black data points). This makes sense from an evolutionary standpoint: slow changes are less likely to signal an immediate threat, so the channels shut off instead of staying alert. The red data points are created by the anti-dashpot mechanism, which is obtained by swapping the roles of two states (inactivated $\leftrightarrow$ open ) in Fig. \ref{F6}B. This hypothetical channel, unlike MscS, completely ignores rapid changes, hence, it would be a terrible emergence valve.}
    \label{F7}
\end{figure}
To understand how MscS helps \textit{E. coli} survive, we first should note that a wild-type \textit{E. coli} is armored with dozens of MscS and MscL channels. These channels kick in at different levels of osmotic stress. As shown in Fig. \ref{F2}B, MscL only activates when all MscS channels are already fully open. This makes MscL the cell’s last line of defense---a true emergency valve. MscS, on the other hand, activates at much lower membrane tension and even if it can single-handedly rescue cells from bursting, it is not really an emergency valve like MscL. What sets MscS apart is its ability to enter a third state known as the \emph{inactivated} state, where it becomes non-conductive and tension-insensitive. This is quite different from the closed state. In the closed state, MscS remains responsive to tension and can open, whereas in the inactivated state, it becomes completely insensitive to tension. Importantly, this new state gives MscS more information-processing capabilities compared to MscL (see also Fig. \ref{F6}A). The topology of the state space is depicted in Fig. \ref{F6}B and all the parameters governing the transition between the states have already been experimentally determined. This landscape gives rise to a discrete-space, continuous time Markov process, which we can simulate as well as compare these simulations with experiments. Below, we list two important features of MscS' information processing capabilities.  

First, MscS channels enter the inactivated state only if the tension is neither too low nor too high, as shown in Fig. \ref{F7}A. The sweet spot for inactivation happens when the tension is near the midpoint tension (denoted as $\gamma_{0.5}$), where the probability of the channel being open is 0.5. The panel displays experimentally determined steady-state levels of MscS inactivation after applying constant tension for 120 seconds, represented by the black data points with error bars. The red curve represents the theoretical prediction of inactivation obtained from a three-state Markov model of the channel (see \cite{ccetiner2018spatiotemporal} for more details), while the blue line is the normalized dose-response curve from a 1-second ramp protocol. When the tension is too low, the energy landscape is such that the channels prefer to stay closed but they are ready to fire if needed. Similarly, when tension is high---signaling a possible emergency---becoming inactivated would be a bad move, as the channels need to stay open and reduce the pressure. At these two extremes, the energy landscape gives negligible inactivation. However, at around the midpoint tension, the channel becomes flickery. It switches back and forth between open and closed states. Given that the midpoint tension is much smaller than the lytic tension of membranes, this indecisive behavior is risky because it could waste cellular resources. MscS solves this problem by maximizing its inactivation at this midpoint, essentially shutting itself off to avoid unnecessary leakage in a situation where the tension is not dangerous. Inactivation is fully reversible. When the tension drops, the channels naturally return to their lowest energy state---the closed state. From there, they are ready to respond when tension is applied again.

Second, MscS fully responds to rapid changes but ignores slow ones, thanks to its ability to enter the inactivated state. Let us run some simulations using the QUBexpress software (\url{https://qub.mandelics.com}) and the channel parameters given in Fig. \ref{F6}. During the simulation, we increase the membrane tension from zero to a value at which the channel is (almost) guaranteed to open, say $3\ k_BT/nm^2$. We do this at different rates, ranging from a quick 0.1 seconds to a very slow 60 seconds. The black dots in Fig. \ref{F7}B show the fraction of open channels depending on how fast or slow the tension is increased. These results are consistent with previous publications \cite{akitake2005dashpot,ccetiner2018spatiotemporal}. When tension is applied quickly, most channels open by the end of the process. But when tension is applied slowly, over several seconds, something interesting happens: the channels tend to enter an inactivated state instead of opening, giving rise to a decrease in the number of open channels. This behavior is previously called the ``dashpot” mechanism, and it may be important in environmental situations where the sense of urgency is critical for bacteria. That is, slow changes are less likely to signal an immediate threat and MscS can sense it. It is also fun to create an ``anti-dashpot" mechanism by swapping the two states (inactivated $\leftrightarrow$ open in Fig. \ref{F6}B) so that inactivated and open states switch roles, while keeping the transition rates the same. As shown in Fig. \ref{F7}B by the red data points, this hypothetical channel ignores fast stimuli but responds more readily when the membrane tension is increased slowly. Such a channel would be a terrible emergency valve but might work well as a regulator for gradual changes in tension. 

%%%%%%%%%%%%%%%%%%%%%%%%%%%%%%%%%%%%%%%%%%%%%%%%%%%%%%%%%%%%%%%%%%%%%%%%%%%%%%%%%%%%%%%%%%%%%%%%%%%%%%%%%%%%%%%%%%%%%%%%%%%%%%%%%%%%%%%%%%%%%%%%%%%%%%%%%%%%%%%%%%%%%%%%%%%%%%%%%%%%%%%%%%%%%%%%%%%%%%%%%%%%%%%%%%%%%%%%%%%%%%%%%%%%%%%%%%%%%%%%%%%%%%%%%%%%%%%%%%%%%%%%%%%%%%%%%%%%%%%%%%%%%%%%%%%%%%%%%%%%%%%%%%%%%%%%%%%%%%%%%%%%%%%%%%%%%%%%%%%%%%%%%%%%%%%%%%%%%%%%%%%
\section*{Conclusions}
While a true emergency valve, such as MscL, can encode one bit of information—0 for no danger, 1 for danger—MscS channels can represent more than one bit of information because of the additional non-conductive and tension-insensitive state. This allows them to distinguish situations where there may be some danger but not enough to declare a full-blown emergency. Instead of immediately dumping cellular contents, the channels inactivate, giving the cell time for other osmoregulatory mechanisms to kick in. These ``smart" features of MscS are highly sensitive to the parameter set defining the energy landscape of the channel. A small change in these transition rates or the topology of the landscape can easily result in a different information processing capability. MscL, for example, is a single-bit channel that cannot feel the rate of tension application—it just reacts. To deal with such subtleties, an extra state with carefully tuned properties is needed. There are many homologs of MscS channels in eukaryotes, including fission yeast, algae, flagellates, and plants. It would be interesting to compare and contrast the information-processing capabilities of these MscS homologs. How much of the original MscS functionality is preserved in eukaryotes? What new features have been added and what kind of biophysical trade-offs come with them? 

Every patch clamp trace is a stochastic trajectory. Recent advancements in our understanding of the thermodynamics of small systems allow us to calculate the energetics of these nanoscale machines by defining heat and work at the level of a single stochastic trajectory. In small systems, due to the random thermal motion of molecules, heat and work take different values from one realization to the next. However, useful information is encoded in the distributions of these quantities. By examining different driving protocols, we can determine which protocols cause the channels to dissipate the least amount of heat and operate most efficiently. The hypothesis is that the driving protocols that minimize dissipation may be similar to the forces that these channels have evolved to respond to. So far, heat distributions have only been obtained for \textit{E. coli}'s MscS. A key step forward would be to gather similar data for other channels. MscL, for example, would be a good candidate. We would expect that, as a true emergency valve, MscL would dissipate less heat than MscS at fast (millisecond) pulling rates. There are also plenty of eukaryotic channels for which we know almost nothing about their thermodynamic efficiency. It is now time for a deeper exploration of how evolution and physics, together, shape the information-processing capabilities of these fascinating nanoscale machines.
%%%%%%%%%%%%%%%%%%%%%%%%%%%%%%%%%%%%%%%%%%%%%%%%%%%%%%%%%%%%%%%%%%%%%%%%%%%%%%%%%%%%%%%%%%%%%%%%%%%%%%%%%%%%%%%%%%%%%%%%%%%%%%%%%%%%%%%%%%%%%%%%%%%%%%%%%%%%%%%%%%%%%%%%%%%%%%%%%%%%%%%%%%%%%%%%%%%%%%%%%%%%%%%%%%%%%%%%%%%%%%%%%%%%%%%%%%%%%%%%%%%%%%%%%%%%%%%%%%%%%%%%%%%%%%%%%%%%%%%%%%%%%%%%%%%%%%%%%%%%%%%%%%%%%%%%%%%%%%%%%%%%%%%%%%%%%%%%%%%%%%%%%%%%%%%%%%%%%%%%%%%%%%
\section*{ACKNOWLEDGMENTS}
U.C. gratefully acknowledges the support of the Maryland Biophysics Program and the U.S. Department of Education GAANN Fellowship in Mathematics in Biology. Special thanks to Sergei Sukharev for valuable feedback and to Jeremy Gunawardena for ongoing support and guidance.
\bibliography{refs}
\end{document}